\documentclass[10pt,aps,prb,showpacs,preprintnumbers,twocolumn,superscriptaddress]{revtex4-2}
\usepackage{amsmath,amssymb}
\usepackage{bm}
\usepackage{tipa}
\usepackage{upgreek}
\usepackage{comment}
\usepackage{mathrsfs}
\usepackage{graphicx}
\usepackage{braket}
\usepackage{enumitem}
\usepackage{mathbbol}
\usepackage{booktabs}
\usepackage{multirow}
\usepackage{textcomp}
\usepackage[normalem]{ulem}
\usepackage{color}

\usepackage{pifont}

\renewcommand{\comment}[1]{}

\usepackage[colorlinks,bookmarks=true,citecolor=blue,linkcolor=red,urlcolor=blue]{hyperref}
\begin{document}

\title{Full, three-quarter, half and quarter Wigner crystals in Bernal bilayer graphene}
\author{Enrique Aguilar-Méndez}
\affiliation{Institute for Theoretical Physics, ETH Z{\"u}rich, 8093 Z{\"u}rich, Switzerland}
\author{Titus Neupert}
\affiliation{Department of Physics, University of Z{\"u}rich, Winterthurerstrasse 190, 8057 Z{\"u}rich, Switzerland}
\author{Glenn Wagner}
\affiliation{Institute for Theoretical Physics, ETH Z{\"u}rich, 8093 Z{\"u}rich, Switzerland}

\begin{abstract}
Application of a displacement field opens a gap and enhances the Van-Hove singularities in the band structure of Bernal-stacked bilayer graphene. By adjusting the carrier density so that the Fermi energy lies in the vicinity of these singularities, recent experiments observe a plethora of highly correlated electronic phases including isospin polarized phases and high-resistance states with non-linear electric transport indicative of a possible Wigner crystal. We perform Hartree-Fock calculations incorporating long-range Coulomb interactions and allowing for translational and rotational symmetry breaking. We obtain the displacement field vs.~carrier density phase diagram which shows isospin polarized metallic phases tracking the Van-Hove singularity in the valence band. Between these metallic phases we observe regions where the ground state is a Wigner crystal. The isospin polarization of the Wigner crystals tracks the isospin polarization of the nearby metallic phases. Depending on whether we have four, three, two or one isospin flavours, we obtain a full, three-quarter, half or quarter Wigner crystal.  
\end{abstract}

\maketitle

\section{Introduction}
\label{sec:introduction}
The recent observation of superconductivity and other exotic phases in Bernal-stacked bilayer graphene (BBG) \cite{Zhou2022,Zhang2023,Holleis2025} has revived interest in this relatively simple graphene allotrope.  These observations occurred in the context of numerous previous discoveries of strongly correlated electronic phases in twisted multilayer graphene (TMG) \cite{Cao2018,Stepanov2020,Oh2021,Yankowitz2019,Park2021,Park2022,Zhang2022,Su2023,Uri2023} and in untwisted trilayer \cite{Zhou2021,Patterson2025}, tetralayer and pentalayer graphene \cite{Han2025}. A commonality is the presence of flat regions in the band structure and corresponding Van Hove singularities. When the Fermi energy is set to the vicinity of these flat regions via electrostatic doping,\comment{the kinetic energy is quenched and} the electronic interaction becomes preponderant. Bearing in mind a Stoner model of ferromagnetism generalized to valley and spin, the presence of phases that break isospin symmetries in these systems \cite{Lu2019,Yu2022,Cao2021,Liu2022,Sharpe2019,Serlin2020,Saito2020,Zondiner2020,Stepanov2021,Wu2021,Pierce2021,Das2021,Saito2021,Seiler2024a}
\comment{last one is BBG with SOC} comes as a result of the exchange energy benefiting from polarization in the valley and spin degrees of freedom. In this way it is possible to obtain metallic states where the charge carriers are not of all four possible isospins (full metal), but instead are polarized into three, two or one isospins. These are denoted as three-quarter, half, and quarter metal, respectively. The Femi surface of these states has a number of pockets equal to the number of isospin species present.

In BBG an out-of-plane displacement field creates a gap in the band structure. The valence and conduction band develop a flat region at the corners of the Brillouin zone. In this work, we focus on the regions proximate to a Van-Hove singularity in the valence band. This singularity can be understood, within a single particle picture, as a transition between a phase with twelve and a phase with four Fermi surface pockets. Magnetoresistance measurements are a way to probe the topology of the Fermi surface. When the resulting quantum oscillation frequencies are normalized by carrier density, the fraction of the Fermi surface enclosed by a cyclotron orbit is obtained. Using this technique for a range of displacement fields and carrier densities, a phase diagram is obtained~\cite{Zhou2022,delaBarrera2022,Seiler2022} that shows additional intermediate phases with different numbers of Fermi surface pockets. This is indicative of isospin polarization or Fermi surface reconstruction due to inter-valley coherence (IVC). 

\begin{figure}[h!]
    \centering
    \includegraphics[width=8cm]{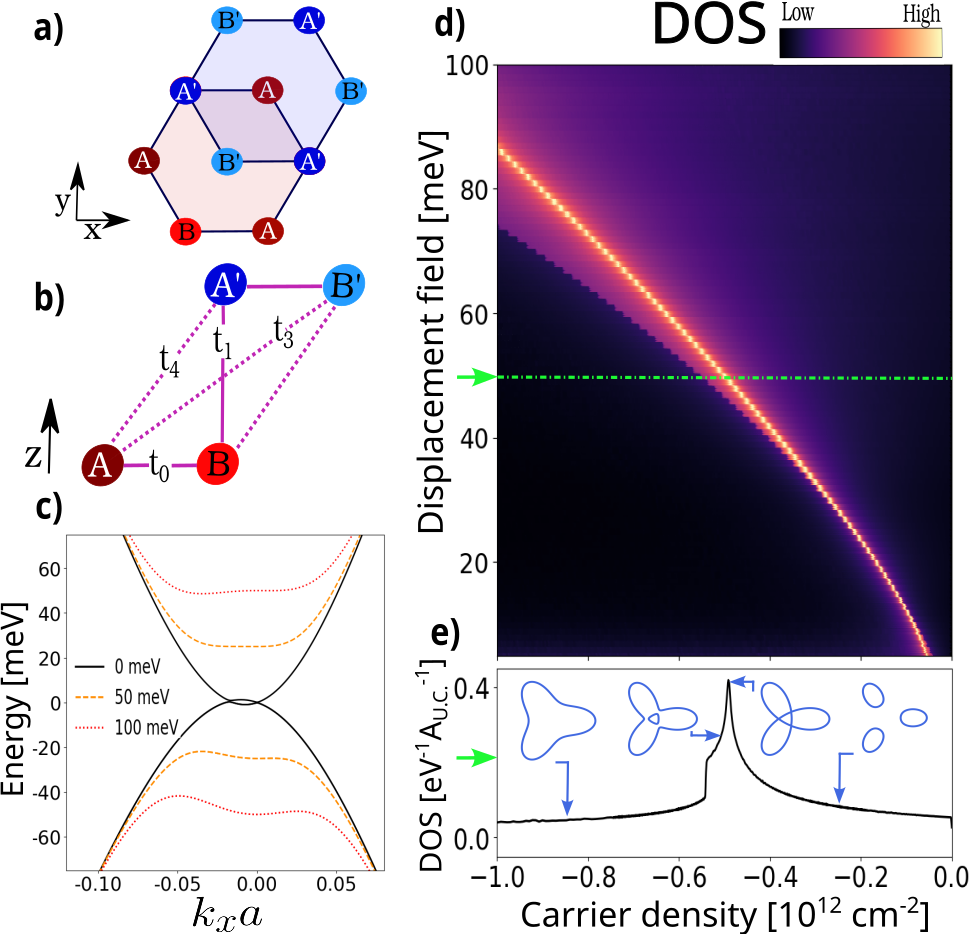}
    \caption{ \textbf{a)} Arrangement of top (blue) and bottom (red) layers in Bernal bilayer graphene. \textbf{b)} Transverse view of an $A'$ atom stacked on top of a $B$ atom. 
    \textbf{c)} Cut along $k_y=0$ of the valence and conduction bands for different displacement fields. The displacement field opens a gap and generates a flat region close to the band edges. The $\bm{K}$ point is located at $k_x=0$. \textbf{d)} Density of states. The Van-Hove singularity moves to larger carrier densities for higher displacement fields. The green arrow indicates a line cut shown in figure e). \textbf{e)} Density of states for a displacement field of 50 meV. The shape of the Fermi surface for a single valley and spin is indicated at 4 different values of carrier density. The Van-Hove singularity corresponds to a Lifshitz transition whereby three pockets in the Fermi surface merge.}
    \label{fig:single_particle}
    \vspace*{-.3cm}
\end{figure}

Close to these intermediate phases, Refs.~\onlinecite{Zhou2022, Seiler2024b} observe regions where the electrical resistance is high and shows nonlinear behavior resembling the depinning process in a charge density wave \cite{Fleming1979}. In Ref.~\onlinecite{Zhou2022} one such state is reported as the possible parent state at zero magnetic field from which superconductivity develops once an in-plane magnetic field is applied. Different mechanisms have been proposed to explain this superconducting state, ranging from phonon-mediated \cite{Chou2022,Chou2022a,Chou2022b} to purely electronic \cite{Pantaleon2023,Dong2023a,Szabo2022,Jimeno-Pozo2023,Cea2023}, however, no consensus has been reached. 
Recently proposed mechanisms incorporate the requirement of either in-plane magnetic field or spin orbit coupling \cite{Curtis2023,Dong2024a,Wagner2024,Friedlan2025}. 

In TMG translational symmetry breaking phases have been predicted \cite{Kwan2021,Wagner2022,Wang2024} and observed \cite{Kim2022,Nuckolls2023}. Graphene multilayers with larger numbers of layers such as pentalayer graphene may also exhibit translational symmetry breaking phases. As these phases simultaneously acquire a Chern number, they are called anomalous Hall crystals \cite{Lu2024,Dong2024c,Dong2024d,Zhou2024,kwan2023moirefractionalcherninsulators,Bernevig2025}.  It is therefore sensible to speculate that BBG can host insulating phases with broken translational invariance. 

In a Wigner crystal (WC), charge carriers arrange in a periodic lattice in order to minimize the Coulomb interaction \cite{Wigner1934}.  For two-dimensional electron gases, the emerging lattice is usually triangular \cite{Attaccalite2002}.
WCs are more stable in the regime of low carrier density given that in this regime Coulomb interactions dominate over kinetic energy \cite{Tanatar1989}. These conditions are met in Landau-level systems at low filling fractions. In these systems WC formation out of a two-dimensional electron gas has been extensively reported \cite{Andrei1988,Jiang1990,Goldman1990,Liu2016,Chen2023}. It competes with other states typical of quantum Hall systems such as the composite Fermi liquid \cite{Amet2015} and both integer \cite{Novoselov2006} and fractional quantum Hall states \cite{Dean2011}. 

There is solid evidence of WCs at zero magnetic field in monolayer \cite{Smolenski2021}, bilayer \cite{Xiang2024} 
and moiré \cite{Xu2020,Regan2020,Huang2021,Li2021,Li2024a,Xiang2024} transition metal dichalcogenides. In BBG a WC has been directly observed only 
in presence of a strong ($\sim$ 14 T)  
magnetic field \cite{Tsui2024}, this was accompanied by the observation of coexisting stripe phases. However, some recent experiments suggest that a WC can occur in BBG at zero magnetic field \cite{Seiler2024b, Seiler2024}. In these experiments, conductance fluctuations are observed above a threshold value of a bias current. Additionally, when an AC current is superimposed on a DC current, current modulations are observed with a frequency that increases linearly with the DC current. These results are observed within a specific range of carrier densities and are consistent with the depinning and sliding processes of a WC. The appearance of phases usually associated with Landau level physics at zero magnetic field in multilayer graphene systems \cite{Lu2024,Lu2025} does not come as a surprise since the flat regions in the band structure can play a similar role as the flat Landau levels.



Analytical methods have been used to explore the possibility of WCs \cite{Silvestrov2017,Joy2023}, charge density waves, and spin density waves \cite{Dong2024} in BBG. Hartree-Fock calculations
have been used on a Hubbard model to explore the latter kind of state \cite{Scholle2024}.


In this work we perform self-consistent Hartree-Fock calculations allowing for translational symmetry breaking. \comment{ ,i.e., allowing the occupation of states that are non-diagonal in momentum} Our starting point is a momentum space tight binding Hamiltonian  on top of which we include long range Coulomb interactions. We obtain displacement field vs. carrier density phase diagrams. 
In regions for which the momentum-non-diagonal contribution of the density matrix is negligible they agree with previous analytical results \cite{Dong2023,Mayrhofer2025} and with Hartree-Fock studies where translational invariance violations were not allowed \cite{Xie2023,Wang2024a, Koh2024}. We observe regions with significant momentum-non-diagonal contribution and large energy gap associated with WC formation in a range of carrier densities and displacement fields consistent with experimental results \cite{Zhou2022,Seiler2024b}. 

\section{Method}
\label{sec:method}

The baseline for our calculations is a nearest neighbors tight-binding Hamiltonian tailored to describe the band structure of BBG around the $\bm{K}$ and $\bm{K'}$ points of the Brillouin zone \cite{Jung2014}:

\begin{equation}
    \label{eq:tight_binding_hamiltonian}
H_{\tau}(\bm{k})=
\begin{pmatrix}
  D/2 & v_0 \pi^{\dagger} & -v_4 \pi^{\dagger} & -v_3 \pi\\ 
  v_0 \pi & D/2+\Delta' &t_1 & -v_4 \pi^{\dagger}\\
  -v_4 \pi& t_1 & -D/2+\Delta' & v_0\pi^{\dagger}\\
  -v_3 \pi^{\dagger}& -v_4 \pi & v_0 \pi & -D/2
\end{pmatrix}, 
\end{equation}
here $\pi\equiv \hbar (\tau k_x+i k_y)$ and $v_i \equiv t_i\sqrt{3}a/2\hbar$. The hopping parameters $t_0 ,t_1, t_3$ and $t_4$ have values 2.61 eV, 0.361 eV, 0.283 eV and 0.138 eV, respectively \cite{Jung2014}. 
 $a$ is the lattice constant of graphene. $D$ is the out-of-plane displacement field. $\Delta'=0.015$ eV accounts for the difference in the onsite energy of different sublattices. See Fig.~\ref{fig:single_particle}. This Hamiltonian operates on the basis $\{A',B',A,B\}$, where an apostrophe is used to differentiate the A and B sublattices of the top and bottom graphene layers.  The index $\tau \in \{+1,-1\}$ corresponds to valleys $\bm{K}$ and $\bm{K'}$ respectively. 

The spin degree of freedom enters the model by employing two copies of the system, each of which has a single particle Hamiltonian given by Eq.~\eqref{eq:tight_binding_hamiltonian}. The system has $SU(2)$ spin symmetry. In order to reduce the high computational cost associated with searching for ground states without translational symmetry and given that spin textures in this material have already been explored \cite{Scholle2024}, we assume spin colinearity and study spin-diagonal states in this work. 
 We only consider states in the highest energy valence band in this four-band Hamiltonian as we are interested in hole doping close to charge neutrality. 

Interactions are introduced through a double-gate screened Coulomb potential \cite{Huang2023}:   \comment{$q=\vert \bm{k}_{a'}-\bm{k}_a \vert$}

\begin{equation}
V(q)=\frac{e^2}{2\epsilon_0\epsilon_rq}\tanh{qd_{\textrm{sc}}} , 
\label{eq:interaction}
\end{equation}
 where $\epsilon_0$ is the vacuum's permittivity, $d_{\textrm{sc}}$  is the gate screening length and $\epsilon_r$ is the relative permittivity. We take $d_{\textrm{sc}}
=25$ nm  and  $\epsilon_r=5$  within the experimentally relevant range \cite{Zhou2022,Laturia2018,Bessler2019}. 

Given that the tight-binding parameters are obtained by benchmarking with density functional theory calculations, they already incorporate some effect of Coulomb interactions. In order to avoid double counting, a reference state is subtracted from the density matrix when the Hartree-Fock Hamiltonian 
is computed as explained in the appendix~\ref{appx:Hartree-Fock and optimal damping algorithm}. There are various options for this reference state used in literature, in this work we use a state corresponding to completely filling the valence band with electrons. This has resulted in accurate results in previous works \cite{Koh2024,Kwan2021}. 
An additional reference projector was also used, which led to qualitatively consistent results (Figs.~\ref{fig:multiple_reference_invariant} and \ref{fig:multiple_reference_WC}).

For each value of displacement field and carrier density we employ a 13 by 13 momentum mesh centered around the $\bm{K}$ and $\bm{K'}$ points. The mesh points are taken within the reduced Brillouin zone  spanned by the reciprocal vectors $G^\mathrm{WC}_1$ and $G^\mathrm{WC}_2$ corresponding to a possible Wigner crystal at such carrier density. The magnitude of the Wigner crystal lattice vector $a_0$ and the carrier density $n$ are related through \cite{Joy2023}
\begin{equation}
a_0=\Big( \frac{2}{\sqrt{3}n} \Big)^{\frac{1}{2}}.
\label{eq:WC_lattice_constant}
\end{equation}

We fold the points in momentum space into the reduced Brillouin zone and truncate at 9 bands per valley and spin, corresponding to a cutoff in momentum space. Namely, if $\bm{p}_{(0,0)}$ denotes a point inside the reduced Brillouin zone, the 9 points obtained through $\bm{p}_{(i,j)}=\bm{p}_{(0,0)}+iG^\mathrm{WC}_1+jG^\mathrm{WC}_2$ for $i,j \in \{-1,0,1\}$ are included in the system. This means a total of $1521$ points in momentum space\comment{are used in the calculation}.  All points obtained for a fixed pair $(i,j)$ are labeled with the same band index. Our calculation captures the possibility of translational invariance violation by allowing electronic ground states with non-diagonal contribution in band index. The optimal damping algorithm \cite{Cances2000,Yamamoto2014} was used to stabilize convergence. 

Within this band-folding scheme in momentum space, we can explore the formation of Wigner crystals by completely filling one to four bands. This corresponds to having holes with one to four isospin flavors per momentum in the reduced Brillouin zone. These four possibilities encompass all the possible isospin polarization scenarios for a WC ground state, which we denote as quarter WC (Q-WC), half WC (H-WC), three-quarter WC (TQ-WC) and full WC (F-WC).
When $N$ bands are full the carrier density per isospin is $n=\frac{n_e}{N}$. This fraction is the density inserted in equation \eqref{eq:WC_lattice_constant} to compute the WC lattice constant. 

Since the norm of the reciprocal vectors $G^\mathrm{WC}_{1,2}$ depends on the carrier density, the spacing between points in the momentum space mesh goes from 0.013 [1/$a$] at a carrier density of 
$-1.0 \times 10^{12}~\mathrm{cm}^{-2}$ when one single band is full to 0.002 [1/$a$] at a carrier density of $-0.1 \times 10^{12}~\mathrm{cm}^{-2}$ when four bands are full.

As a point of comparison, calculations were also performed on a system in which translational symmetry was imposed. In this case, the band folding procedure within the reduced Brillouin zone was not necessary. This more conventional Hartree-Fock calculations were performed on a triangular mesh of 1801 points in momentum space centered around the $\bm{K}$ and $\bm{K'}$  points. The points were taken in an hexagon of side 0.12 [1/$a$], corresponding to a cutoff in momentum space. The distance between neighboring points is 0.005 [1/$a$].

\section{Results}


\label{sec:results}

\begin{figure}
    \centering
    \includegraphics[width=8cm]{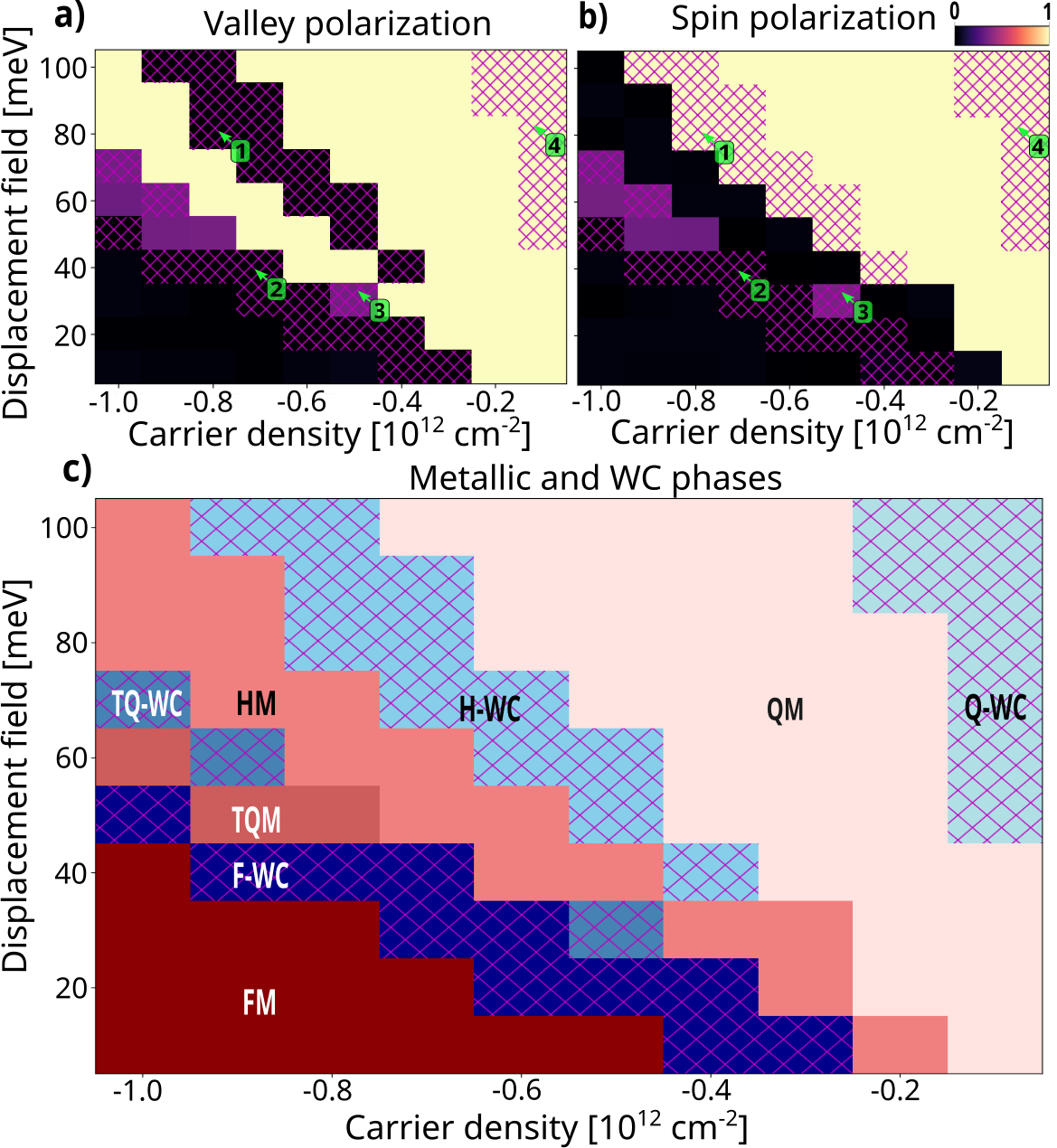}
     \caption{ 
    \textbf{a)} Phase diagram for valley polarization. The solid colors indicate the momentum-diagonal contributions of the ground state's density matrix. The magenta criss-cross hatching indicates regions with large total momentum-non-diagonal contribution. 
    The real space profile of the numbered points is shown in Fig.~\ref{fig:Real_space}. 
    \textbf{b)} Same as a) for spin polarization. \textbf{c)} Figure of all the phases present. The full metal (FM), three-quarter metal (TQM), half metal (HM) and quarter metal (QM) are indicated in shades of red. The full WC (F-WC), half WC (H-WC), three-quarter WC (TQ-WC) and quarter WC (Q-WC) are indicated in shades of blue.  
    }
    \label{fig:phase_diagrams}
    \vspace*{-0.5 cm}
\end{figure}

For each value of carrier density and displacement field in the phase diagrams in figure~\ref{fig:phase_diagrams}, the lowest energy state among the possibilities of one to four bands in the reduced Brillouin zone being full is shown. 
The comparison of the energy per hole for these four scenarios is shown in Fig.~\ref{fig:energy_1_to_4_bands} for a line cut at 50 meV. A list of the order parameters and symmetries of all observed phases is included in appendix \ref{appx:Order parameters and symmetries of observed phases}.

These phase diagrams show a full metal, three-quarter metal, half metal and quarter metal. In the half metal region the ground states can be spin-polarized, valley-polarized or spin-valley-locked. These three states are degenerate 
due to the $SU(2)\times SU(2)$ spin symmetry which allows independent spin rotations in each valley, 
and the time-reversal invariance that connects both valleys. In Fig.~\ref{fig:phase_diagrams} only the valley polarized state is shown for clarity.

We neglect the short range part of Coulomb interactions (inter-valley Hund’s coupling), its effects on BBG have been explored in Ref.~\onlinecite{Koh2024}. This term reduces the $SU(2)\times SU(2)$ symmetry to a global $SU(2)$ symmetry by making it energetically favorable for spins in opposite valleys to align in the same or opposite direction. This term lifts the degeneracy reducing the energy of the spin polarized state in the former case and that of the spin-valley-locked state in the latter.



We do not observe intermediate IVC phases as reported in Refs.~\onlinecite{Xie2023} and~\onlinecite{Koh2024}. This is expected since here we work on a regime of stronger interactions (smaller $\epsilon_r$) that favors exchange energy over kinetic energy and therefore favors isospin polarization over IVC.



There are four additional regions. In these regions translational invariance is broken. They are indicated with criss-cross hatching. Going from left to right in the phase diagram these regions correspond to a Q-WC, H-WC, TQ-WC and F-WC. For the Q-WC the momentum-non-diagonal contribution of the density matrix comes from the valley and spin polarization channels. For the H-WC, TQ-WC and F-WC the momentum-non-diagonal contribution comes from the IVC channel.

The Q-WC, H-WC and F-WC have an energy gap. The TQ-WC is gapless since for this phase holes with two isospins, out of the three present, form a WC in one single spin channel. There is a gap in energy in this spin channel. The holes in the other spin channel remain metallic.

\begin{figure}
    \hspace*{-0.3 cm}
    \centering
    \includegraphics[width=8cm]{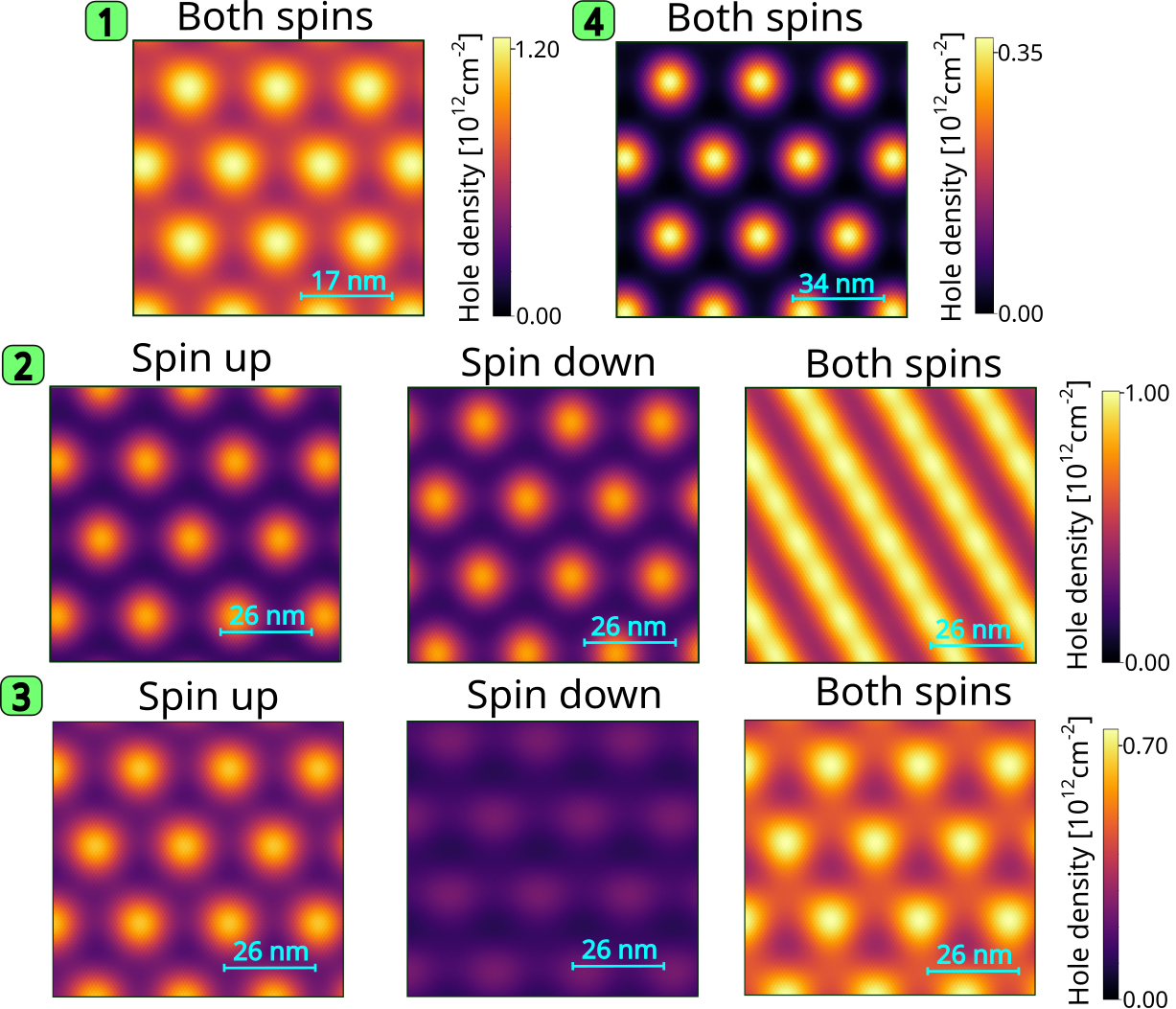}
    \caption{ \textbf{a)} Real space profile of the ground state at different points in the phase diagram (indicated in Fig. \ref{fig:phase_diagrams}). 1 and 4 correspond to a half WC  and quarter WC, respectively. Both are fully spin polarized. 2 and 3 correspond to a full WC and a three-quarter WC, respectively. For these two latter profiles the real space profiles are indicated in each spin channel as well as the pattern resulting from the addition of both profiles.}
    \label{fig:Real_space}
    \vspace*{-0.5 cm}
\end{figure}

In the full WC 
one WC is formed per spin. The addition of both WCs leads to different real space profiles such as the charge-density-wave-like profile shown in panel 2 of Fig.~\ref{fig:Real_space}. Here, the WC profile in each spin and their addition break $C_3$ symmetry. The other panels in this figure show real space profiles of some of the Q-WC, H-WC and TW-WC ground states. All of these profiles have periodicity of a few tens of nanometers. Scanning tunneling microscopy techniques have already been developed that can be used to directly verify their existence \cite{Tsui2024}.

\begin{figure}[t]
    \hspace*{-0.1 cm}
    \centering
    \includegraphics[width=8cm]{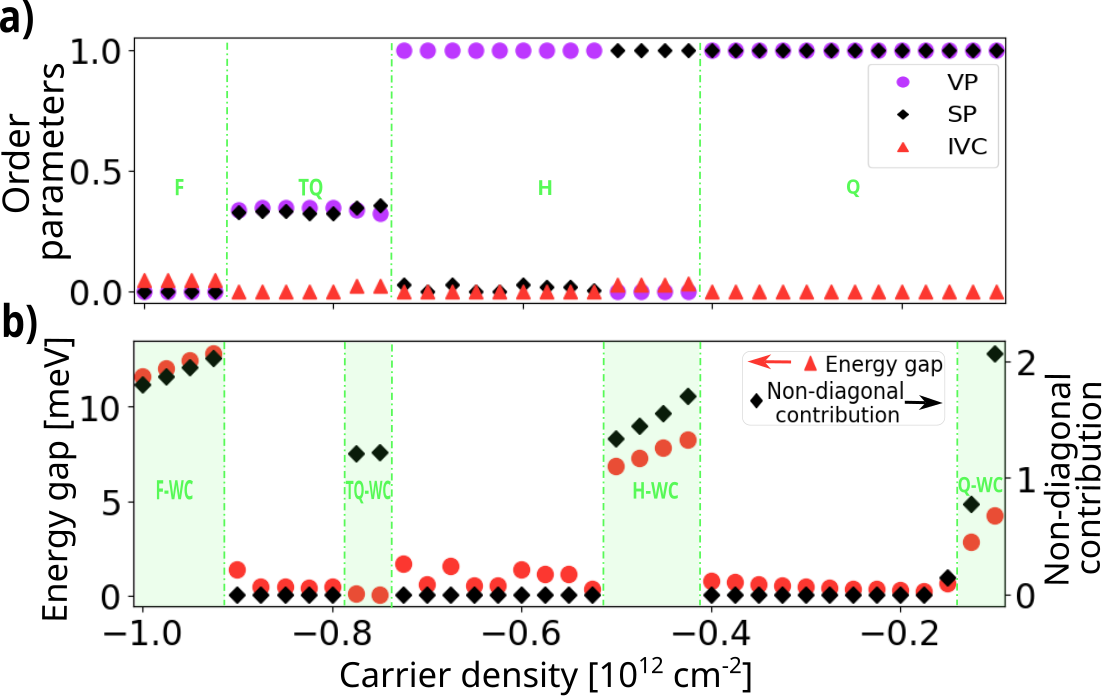}
    \caption{ Line cuts at 50 meV displacement field. \textbf{a)} Momentum-diagonal contribution of the density matrix for valley polarization (VP), spin polarization (SP) and inter valley coherence (IVC). 
    Going from left to right, four regions can be distinguished: a region with no SP and no VP (F), a region with partial VP and SP (TQ), a region with either full SP or full VP (H) and a region with full SP and VP (Q). 
    \textbf{b)} Energy gap (left axis) and total non-diagonal contribution (right axis). Going from left to right, four regions with large non-diagonal contribution can be seen, they correspond to a full WC (F-WC), a three-quarter WC (TQ-WC), a half WC (H-WC) and a quarter WC (Q-WC).
    }
    \label{fig:line_cuts} 
\end{figure}

Figure~\ref{fig:line_cuts} shows a line cut at 50 meV displacement field. In panel a) four regions can be seen in the momentum-diagonal contributions of SP, VP and IVC. At large carrier densities there is an unpolarized full metal region. As the carrier density approaches zero, the interaction strength increases and we observe consecutive transitions to a three-quarter, half and quarter metal. 
Panel b) shows the energy gap and total momentum-non-diagonal contribution of the density matrix. Three regions can be seen where the energy gap is large, going from larger to lower carrier density. These correspond to a full WC, half WC and quarter WC. The approximate energy gaps for these three states are 12 meV, 7 meV and 4 meV, respectively. 
All these energies are considerably larger than the thermal energy corresponding to a few kelvin. The WC phases reported in Ref.~\onlinecite{Seiler2022} lose their characteristic low conductance signature above 10~K. For a WC obtained in presence of a magnetic field, melting is reported above 3~K \cite{Tsui2024}. The WC phases found in this work are expected to be robust in the typical temperature range used in experiments.

In Ref.~\onlinecite{Zhou2022} a high-resistance dome at zero magnetic field is reported from which superconductivity develops once an in-plane magnetic field is applied. At a fixed displacement field, the dome is delimited at larger carrier densities by a half metal region and at smaller carrier densities by a full metal. 
The appearance of a spin unpolarized ground state that breaks translational invariance located in the interface between a isospin-polarized and a isospin-unpolarized phase is consistent with the full WC that we find in this work. The fact that the reported dome extends over a smaller range of carrier densities and appears at a larger displacement field, compared to the full WC in the phase diagram presented here, can be explained by the rather high interaction regime explored in this work. For a higher value of the dielectric constant $\epsilon_r$ the WC phases are expected to shrink towards regions of the phase diagram with higher displacement field where the peak in DOS is most strongly enhanced.  In a regime of weaker electron-electron interaction a full metal region located close to zero carrier density, which is missing in Fig.~\ref{fig:phase_diagrams}, is also expected. 

In Refs.~\onlinecite{Seiler2022, Seiler2024b} a spin polarized WC is reported, which is delimited at lower carrier densities by a half metal and at larger carrier densities by a quarter metal. This is consistent with the H-WC found here. 

The results for a system with enforced translational invariance are shown in Fig.~\ref{fig:multiple_reference_invariant}. Close to charge neutrality we observe a phase with broken $C_3$ rotational symmetry.  The Q-WC is located roughly in the same region of the phase diagram in Fig.~\ref{fig:phase_diagrams}. This close energy competition between a WC and a phase with broken rotational symmetry may be behind the high resistance state reported in Ref.~\onlinecite{Holleis2025}. This state has non-linear resistance behavior and competes with a superconducting phase whose parent normal state is nematic.

\section{Conclusion}
\label{sec:conclusion}
\comment{in \cite{Tsui2024} they observe WC, distorted WC and stripe phases very close in the phase diagram}
\comment{Paper "Wigner crystal phases in bilayer graphene" predicts two WC with different lattice constant in the conduction band at small doping.}

By allowing the ground state of BBG to break translational, rotational, and isospin symmetries, we obtained its 
phase diagram within a Hartree-Fock framework revealing additional phases next to the isospin-polarized full, three-quarter, half, and quarter metals. The new phases break translational symmetry, resulting in different WC profiles. Such phases exhibit reduced electrical conductivity due to the presence of an energy gap in one or both spin channels.

One of these newly identified phases is 
located in the interface between an isospin-polarized and an isospin-unpolarized \comment{translational invariant}metallic phase. This result suggests that the experimentally observed high resistance state from which superconductivity develops in the presence of an in-plane magnetic field (Ref.~\onlinecite{Zhou2022}) can be a WC. 
A mechanism whereby superconductivity emerges from a polarized WC has been proposed for rhombohedral tetralayer graphene \cite{Dong2025}. Our results invite to explore the existence of a similar mechanism in bilayer graphene. Recent work suggests a tendency for two-dimensional Wigner crystals to self-dope in order to become metallic \cite{Kim2024Wigner}, thus forming a parent state from which superconductivity could potentially emerge. 




We also observe a spin polarized WC, which agrees with experimental observations of a spin polarized WC located between a half metal and a quarter metal (Refs.~\onlinecite{Seiler2022, Seiler2024b}) near the Van-Hove singularity.

If we allow only for violation of rotational symmetry, we observe a nematic region close to charge neutrality. This region is superseded by a spin- and valley-polarized WC when translational symmetry breaking is allowed. The close energy competition between these two phases is reminiscent of the competition between a high-resistance state and a superconducting phase that emerges from a nematic normal state in BBG deposited on a WSe$_2$ substrate \cite{Holleis2025}.

\begin{acknowledgments}
\textit{Acknowledgments}.---E.A.M. and G.W.~are supported by the Swiss National Science Foundation (SNSF) via Ambizione grant number PZ00P2-216183. T.N. acknowledges support from the Swiss National Science Foundation through a Consolidator Grant (iTQC, TMCG-2\_213805). 
\end{acknowledgments}



\bibliographystyle{apsrev4-2}
\bibliography{references}

\newpage
\clearpage
\begin{appendix}
\onecolumngrid
	\begin{center}
		\textbf{\large --- Supplementary Material ---\\Full, three-quarter, half and quarter Wigner crystals in Bernal bilayer graphene}\\
		\medskip
		\text{Enrique Aguilar-Méndez, Titus Neupert and Glenn Wagner}
	\end{center}
	
		\setcounter{equation}{0}
	\setcounter{figure}{0}
	\setcounter{table}{0}
	\setcounter{page}{1}
	\makeatletter
	\renewcommand{\theequation}{S\arabic{equation}}
	\renewcommand{\thefigure}{S\arabic{figure}}
	\renewcommand{\bibnumfmt}[1]{[S#1]}

\section{Hartree-Fock and optimal damping algorithm}
\label{appx:Hartree-Fock and optimal damping algorithm}


Starting from the following interacting Hamiltonian:
\begin{equation}
    H=H^\textrm{sp}_{r,r'} c^{\dagger}_r c_{r'}+\frac{1}{2} V_{r,s,r',s'} c^{\dagger}_r c^{\dagger}_s c_{s'} c_{r'}
    \label{eq:interacting_hamiltonian}
\end{equation}
where $H^\textrm{sp}_{r,r'}$ is the single particle part and $V_{r,s,r',s'}$ the interaction the following Hartree-Fock Hamiltonian is obtained: 
\begin{equation}
H^\textrm{HF}_{s,s'}= H^\textrm{sp}_{s,s'}+\underbrace{(V_{r,s,r',s'}-V_{r,s,s',r'})}_{\equiv H^\textrm{int}_{r,r',s,s'}}P_{r,r'}.
    \label{eq:hami_hartree-fock}
\end{equation}


Here $r, s, r'$ and $s'$ stand for combined momentum, spin and valley indices. $P=\sum_{i\leq N_e} \ket{u_i}\bra{u_i}$ projects over occupied states where $\ket{u_i}$ are the eigenvectors of the Hartree-Fock Hamiltonian. If we denote a reference projector by $P^0$, the energy is:
\begin{equation}
    \label{eq:energy}
E_G=\textrm{tr}(H^\textrm{sp}P)+ \frac{1}{2}\textrm{tr}((P-P^0)^2 H^\textrm{int})
\end{equation}


Fig.~\ref{fig:flow_diagram_program} shows the scheme followed to obtain the ground state using equations \eqref{eq:hami_hartree-fock} and \eqref{eq:energy}:

\begin{figure}[h!]
    \centering
    \includegraphics[height=5cm]{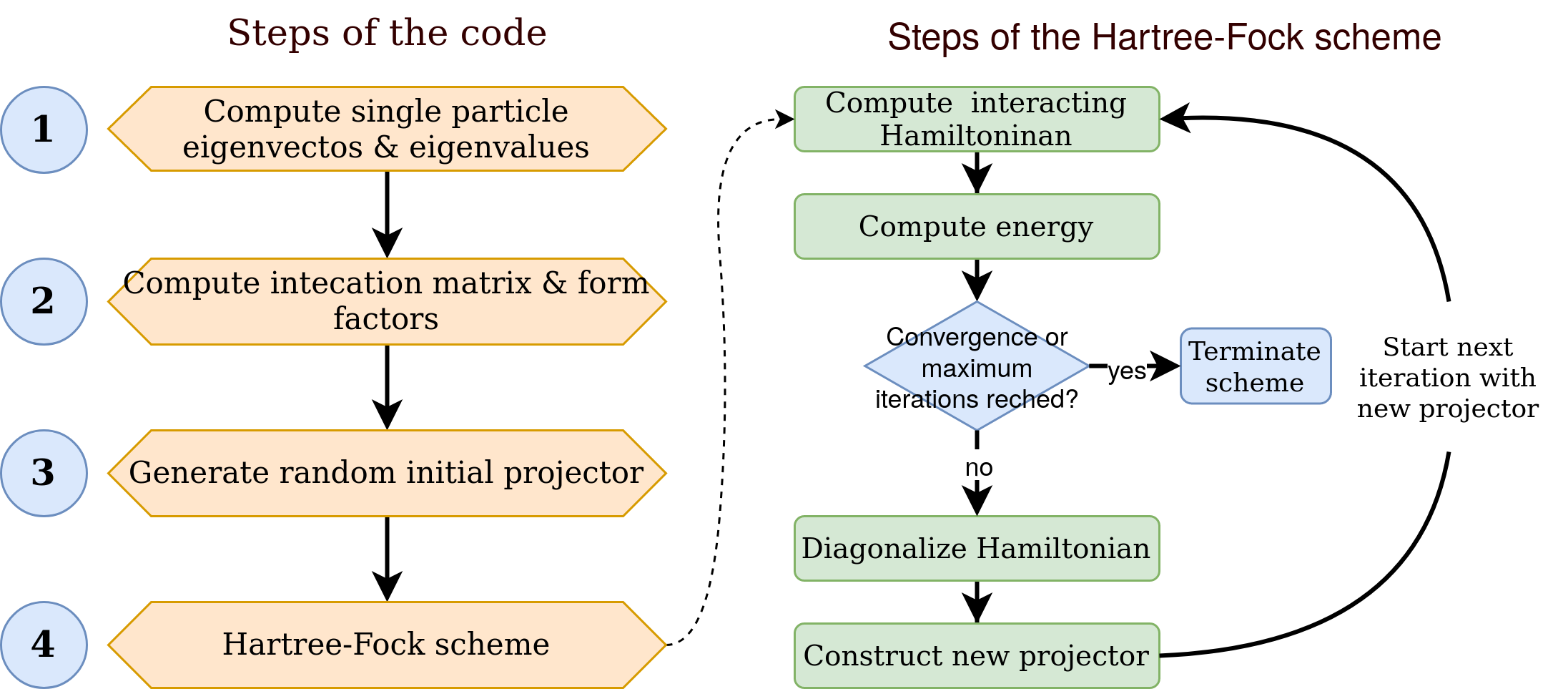}
    \caption{General flow diagram of the self-consistent Hartree-Fock scheme.}
    \label{fig:flow_diagram_program}
\end{figure}

$P^0$ was constructed according to Ref.~\onlinecite{Koh2024}, it represents a state where the valence band has been completely filled with electrons. During the construction of the new projector the optimal damping algorithm (ODA) is used to stabilize convergence. This algorithm interpolates between the projector $P_N$  used to compute the Hartree-Fock Hamiltonian $H^\textrm{HF}_N$ in the $N$'th iteration and the projector $P^\textrm{new}$ generated with the eigenvectors of $H^\textrm{HF}_N$ \cite{Cances2000,Yamamoto2014}. This method produces a projector $P_{N+1}=(1-\lambda)P_N+\lambda P^\textrm{new}$ with $\lambda \in [0,1]$ that will be used as a starting point in iteration $N+1$. $\lambda$ is calculated so that the energy $E^\textrm{HF}_{N+1}$ is minimized. $E^\textrm{HF}_{N+1}$ is quadratic in $\lambda$:  

\begin{equation}
    \label{eq:energy_damping_simple}
\begin{split}
E^\textrm{HF}_{N+1}(\lambda)=&
\textrm{tr}(H^\textrm{sp}P_N)+\frac{1}{2}\textrm{tr}(P_c^2H^\textrm{int})\\ 
&+\lambda \underbrace{(\textrm{tr}(H^\textrm{sp}dP)+\textrm{tr}(P_c dP H^\textrm{int}))}_{C_\textrm{linear}}
+\lambda^2 \underbrace{\frac{1}{2}\textrm{tr}(dP^2 H^\textrm{int})}_{C_\textrm{quadratic}}, 
\end{split}
\end{equation}



where $P_c=P_N-P^0$ and $dP=P^\textrm{new}-P_N$. $E^\textrm{HF}_{N+1}(0)$ is the energy of the system with projector $P_N$  and $E^\textrm{HF}_{N+1}(1)$ is the energy of the system with projector $P^\textrm{new}$. The inflection point of this parabola is: 

\begin{equation}
\label{eq:lambda_optimal}
\begin{split}
\lambda=\frac{-C_\textrm{linear}}{2C_\textrm{quadratic}}. 
\end{split}
\end{equation}

\begin{figure}[h!]
    \centering
    \includegraphics[height=6cm]{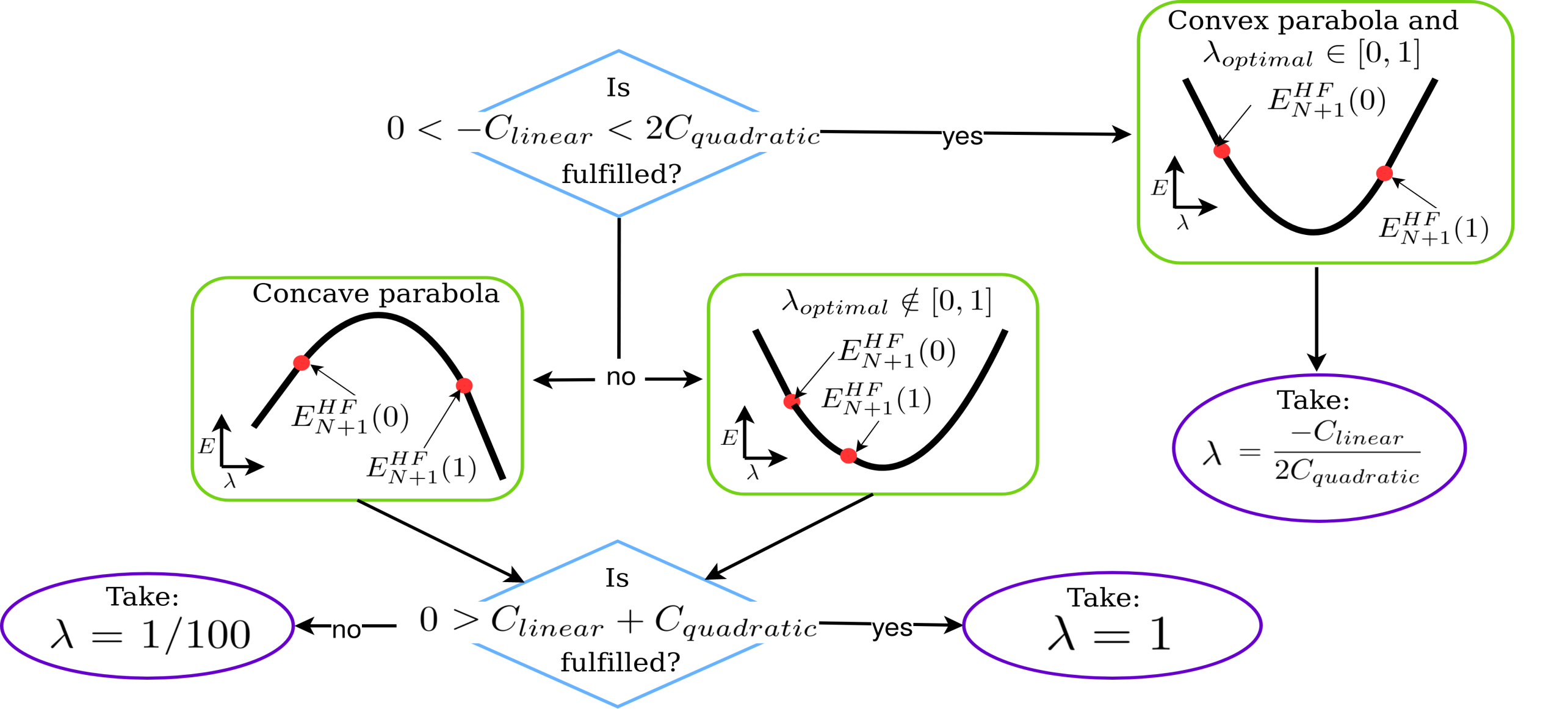}
    \caption{Flow diagram for the optimal damping algorithm.}
    \label{fig:flow_diagram_optimal_damping}
\end{figure}

$\lambda$ corresponds to a minimum in energy only if we are dealing with a convex parabola ($0<C_\textrm{quadratic}$), which is equivalent to: $0<-C_\textrm{linear}<2C_\textrm{quadratic}$. If this is not satisfied we cannot use the value of $\lambda$ given in \eqref{eq:lambda_optimal}. In this case we proceed to check $E^\textrm{HF}_{N+1}(0)>E^\textrm{HF}_{N+1}(1)$, which is equivalent to $0>C_\textrm{linear}+C_\textrm{quadratic}$. If this is satisfied we set $\lambda=1$ meaning that the $P^\textrm{new}$ indeed decreases the energy of the system. Otherwise $P_N$ is energetically favorable and we choose $P_{N+1}$ to be very close to $P_N$, we set $\lambda$ to a small positive number such as $1/100$. Taking $\lambda = 1/100$ instead of $\lambda=0$ allows the code to run for at least a few more iterations thus decreasing the energy of the system a little bit more. Figure Fig.~\ref{fig:flow_diagram_optimal_damping} summarizes this procedure.




\section{Energy of different band fillings in the folded Brillouin zone}

\label{appx:Energy of different band fillings in the folded Brillouin zone}

\begin{figure}[h!]
    \centering
    \includegraphics[height=4cm]{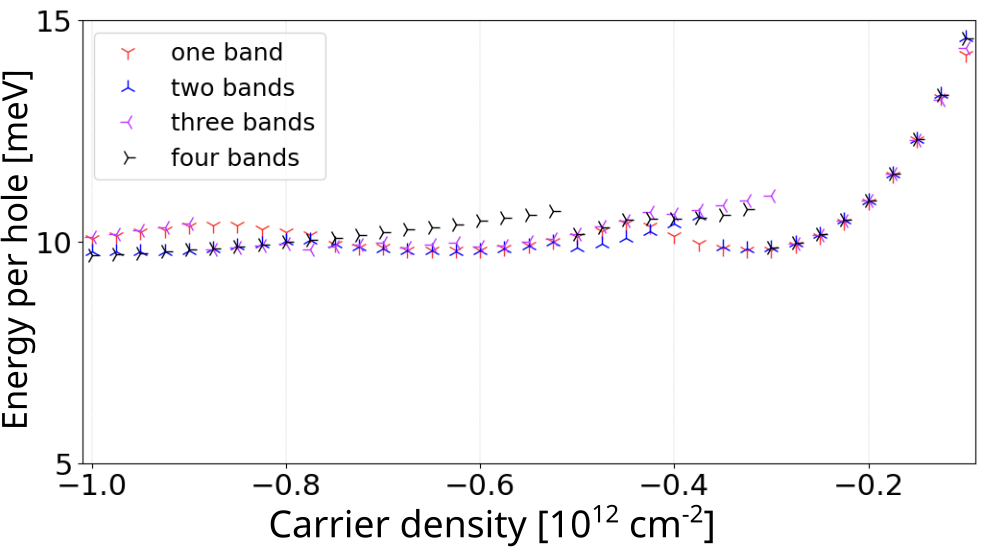}
    \caption{Lowest energy per hole among multiple initial random states for one to four full bands in the reduced Brillouin zone. The displacement field is 50 meV.} 
    \label{fig:energy_1_to_4_bands}
\end{figure}

We study the formation of ground states with broken translational invariance by completely filling one to four bands in the folded Brillouin zone. For each point in the displacement field vs carrier density phase diagram we consider these four possibilities with multiple initial random states. Among all the resulting Hartree-Fock self consistent states the lowest in energy is selected as the ground state. In order to appropriately compare the energy of the states with different number of full bands in the reduced Brillouin zone, we subtract the energy of the corresponding state with a completely full valence band and divide this difference by number of holes. Fig.~\ref{fig:energy_1_to_4_bands} shows, for a line cut at 50 meV, the resulting lowest energy among multiple initial random states for one to four full bands.

\newpage
\section{Order parameters and symmetries of observed phases}
\label{appx:Order parameters and symmetries of observed phases}

\begin{table}[h!]
\vspace{-0.7cm}
\centering
\begin{tabular}{|c|c|c|c|c|c|c|c|}
\hline
\multirow{2}{*}{\centering Phase} & \multirow{2}{*}{\centering VP} & \multirow{2}{*}{\centering SP} & \multirow{2}{*}{\centering IVC} & \multicolumn{4}{c|}{Symmetry} \\ \cline{5-8}
 &  &  &  & $T_R$ 
 & $U_v(1)$ & $SU(2)$ & $C_3$ \\ \hline
FM & $\times$ & $\times$ & $\times$ & $\checkmark$ & $\checkmark$ & $\checkmark$ & $\checkmark$ \\ \hline
F-WC & $\times$ & $\times$ & $\checkmark$ & $\times$ & $\times$ & $\checkmark$ & $\times$ \\ \hline
TQM & $\checkmark$ & $\checkmark$ & $\times$ & $\checkmark$ & $\checkmark$ & $\times$ & $\checkmark$ \\ \hline
TQ-WC& $\checkmark$ & $\checkmark$ & $\checkmark$ & $\times$  & $\times$ & $\times$  & $\times$  \\ \hline
HM & \tiny{$\checkmark|\times|\checkmark$} & \tiny{$\times|\checkmark|\checkmark$} & $\times$ & $\checkmark$ & $\checkmark$ & \tiny{$\checkmark|\times|\times$} & $\checkmark$ \\ \hline
H-WC & $\times$ & $\checkmark$ & $\checkmark$ & $\times$  & $\times$ & $\times$ & $\times$ \\ \hline
QM & $\checkmark$ & $\checkmark$ & $\times$ & $\checkmark$ & $\checkmark$ & $\times$ & $\checkmark$ \\ \hline
Q-WC & $\checkmark$ & $\checkmark$ & $\times$ & $\times$ & $\checkmark$ & $\times$ & $\checkmark$ \\ \hline
\end{tabular}
\caption{Order parameters and symmetries for all phases present in the right column of Fig.~\ref{fig:phase_diagrams}: full metal (FM), full WC (F-WC), thee quarter metal (TQM), three-quarter WC (TQ-WC), half metal (HM), half WC (H-WC), quarter metal (QM) and quarter WC (Q-WC). It is indicated whether valley polarization (VP), spin polarization (SP) and inter valley coherence (IVC) are present or not. It is indicated whether the following symmetries are broken or not: translational invariance ($T_R$), $U_v(1)$ valley charge symmetry and $SU(2)$ spin rotation. Since the Half metal is degenerate in energy: we indicate whether order parameters and symmetries are present for the valley polarized half metal, the spin polarized half metal and the spin-valley-locked half metal. Although the Hamiltonian is spin diagonal, we consider that $SU(2)$ is preserved in the absence of spin polarization. Although the TQ-WC and H-WC shown in Fig.~\ref{fig:Real_space} preserve $C_3$ symmetry, for some carrier densities the triangular WC lattice is slightly deformed and breaks $C_3$ symmetry.}
\label{table:momentum_diag}
\end{table}

\section{Translational invariant results}
\label{appx:translational invariant results}

\begin{figure}[h!]
    \hspace*{-1 cm}
    \centering
    \includegraphics[width=8cm, height=9.5cm]{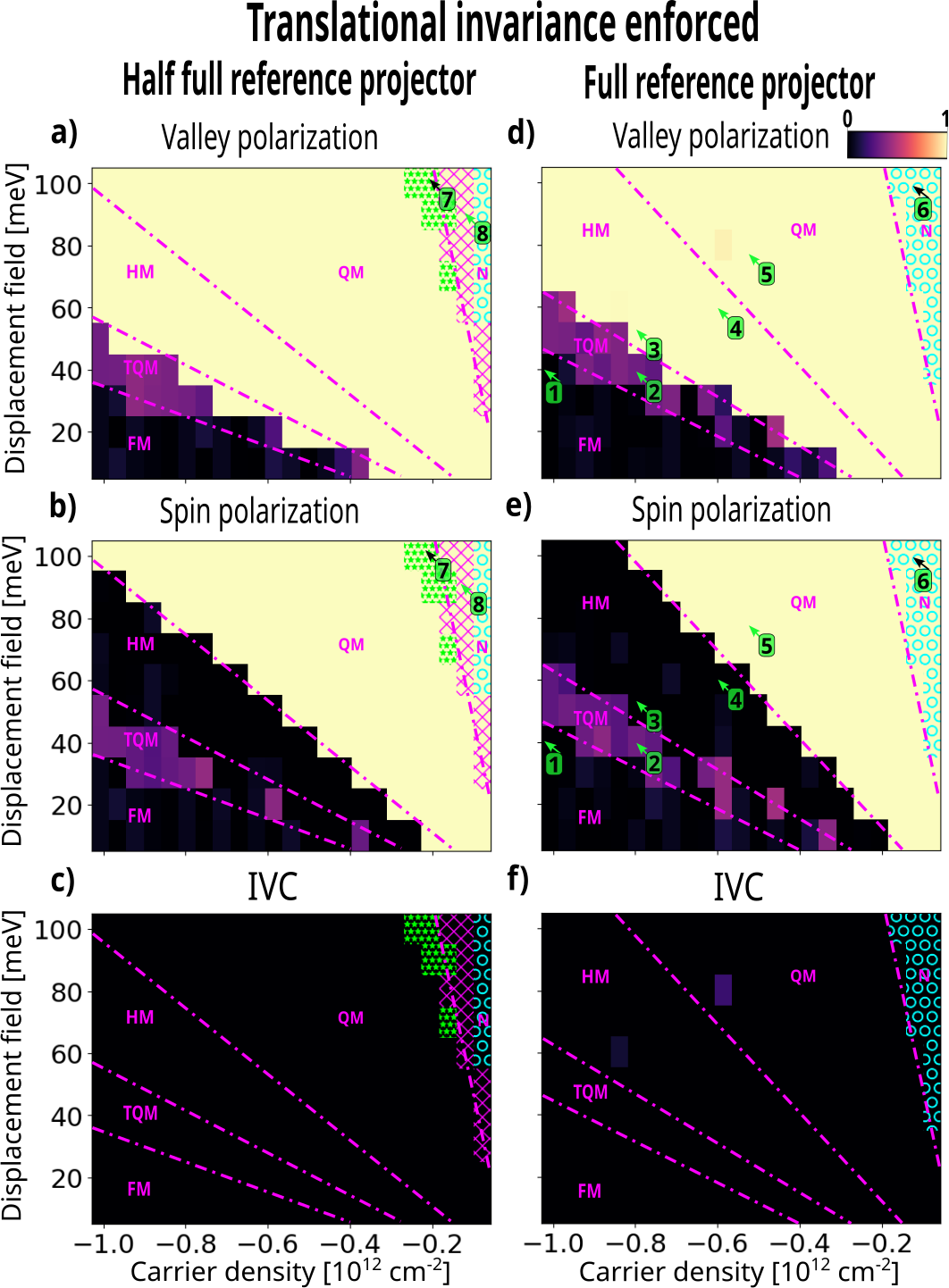}
    \caption{ 
    Valley polarization, spin polarization and inter valley coherence. 
    The full metal (FM), three-quarter metal (TQM), half metal (HM) and quarter metal (QM) regions are indicated. The HM is degenerate and can be valley polarized, spin polarized or spin-valley-locked, only the valley polarized ground state is shown here for clarity. The green starred, magenta criss-crossed, and blue circular hatching indicate the regions where the single spin and valley polarized Fermi surface of the QM breaks into thee, two and one pocket respectively. The former two regions are nematic (N). 
     \textbf{a)}, \textbf{b)}, \textbf{c)} were obtained with a reference projector that corresponds to populating half of the states in the valence band. \textbf{d)}, \textbf{f)}, \textbf{g)} were obtained with a reference projector that corresponds to entirely populating the valence band.
    } 
    \label{fig:multiple_reference_invariant}
\end{figure}

\begin{figure}[h!]
    \hspace*{-0.3 cm}
    \centering
    \includegraphics[width=13cm]{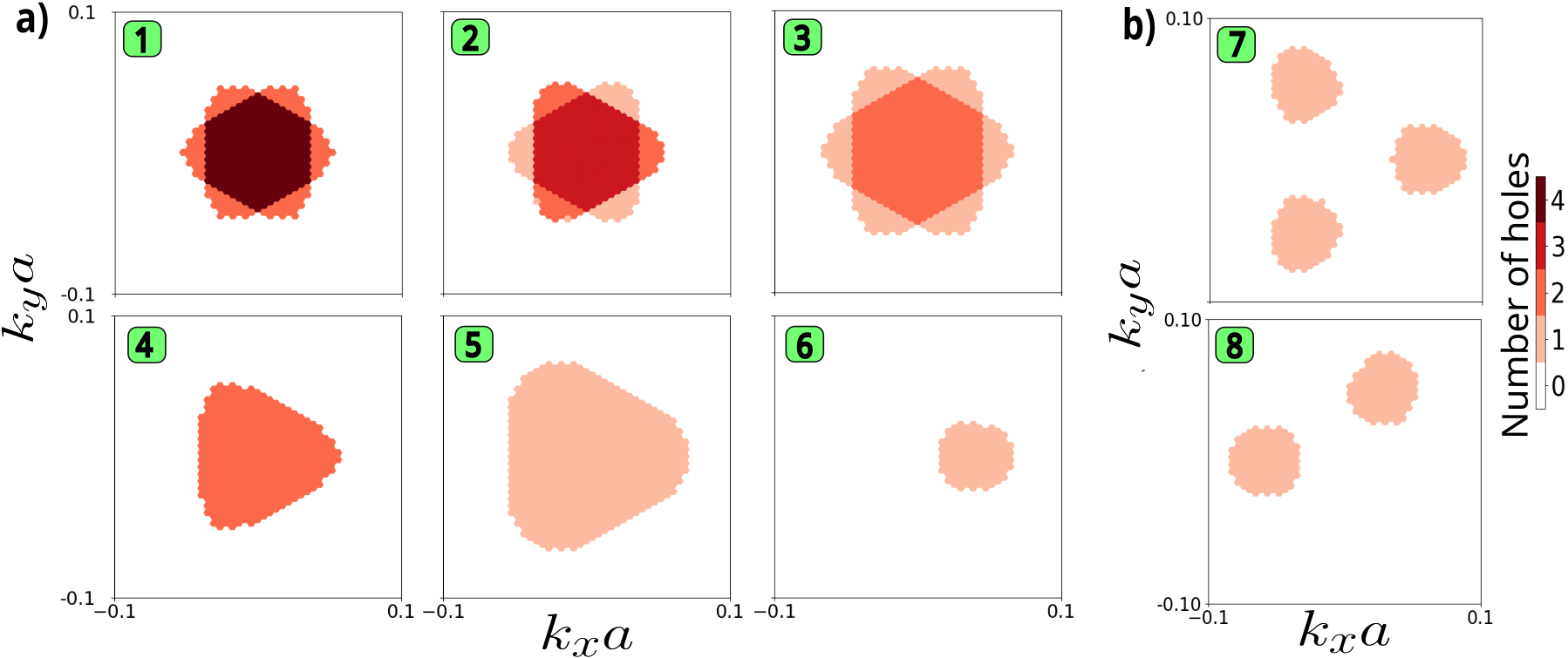}
    \caption{\textbf{a)}, \textbf{b)} Respectively: Fermi surfaces obtained at different points of the phase diagrams in the right and left column of Fig. \ref{fig:multiple_reference_invariant}. 
    Surface 3 corresponds to a spin polarized half metal and surface 4 corresponds to a valley polarized half metal, these two ground states are degenerate in the half metal region, in Fig.~\ref{fig:multiple_reference_invariant} only the valley polarized ground state is shown for clarity. Surfaces 6 and 8 break $C_3$ symmetry.
}
    \label{fig:Fermi_surfaces_and_resl_space}
\end{figure}

Fig.~\ref{fig:multiple_reference_invariant} shows phase diagrams in which translational invariance was enforced. No IVC is observed. Starting from high carrier densities and low displacement fields and moving towards lower carrier densities, four regions can be distinguished corresponding to a full metal, three-quarter metal, half metal and quarter metal. The respective Fermi surfaces of these ground states are shown in Fig.~\ref{fig:Fermi_surfaces_and_resl_space}. 

Close to zero carrier density in the quarter-metal region, the single Fermi surface breaks the $C_3$ rotation symmetry. This region is indicated by circular hatching in Fig.~\ref{fig:phase_diagrams}. The corresponding Fermi surface is marked with number 6 in Fig.~\ref{fig:phase_diagrams}. 
These results are in perfect agreement with Ref.~\onlinecite{Dong2023} where an interaction-induced cascade transition from full metal to quarter metal is predicted as well as the sequential breakdown of the Fermi surface into three, two, and one pockets. 
The breakdown of the Fermi surface into two or one pocket can result in a nematic phase in a similar manner to what has been observed through Hartree-Fock calculations for rhombohedral trilayer \cite{Huang2023} and tetralayer \cite{Parra-Martinez2025} graphene. However, when translational symmetry breaking is allowed, this nematic region is superseded by a WC that preserves $C_3$ rotational symmetry.

\newpage
\clearpage
\section{Comparison of different reference projectors}
\label{appx:Comparison of different reference projectors}

\begin{figure}[h!]
    \centering
     \includegraphics[width=8cm, height=9.5cm]{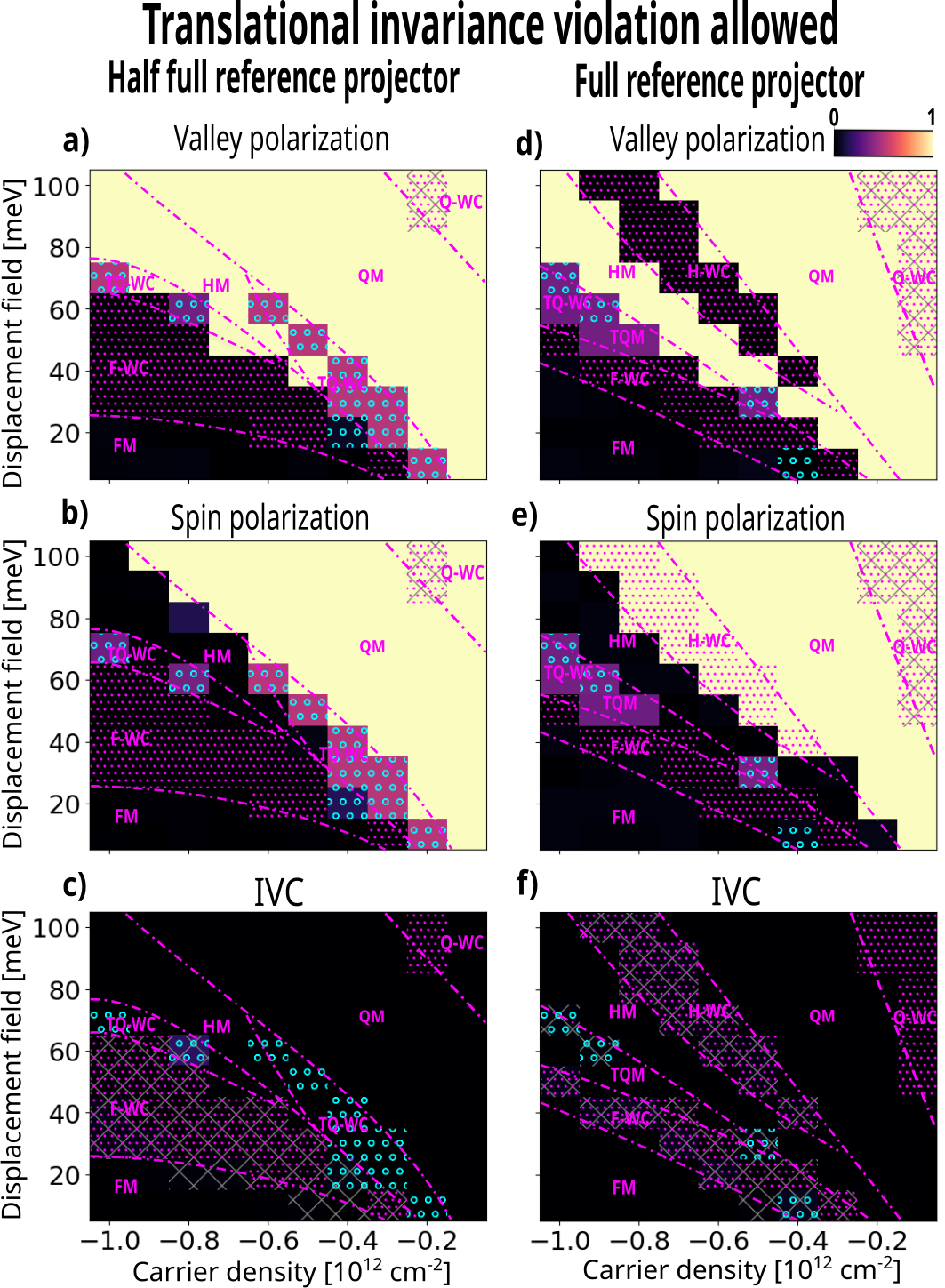}
    \caption{ Momentum-diagonal contribution of the valley polarization, spin polarization and inter valley coherence. 
    The full metal (FM), three-quarter metal (TQM), half metal (HM) and quarter metal (QM) regions are indicated. The HM is degenerate and can be valley polarized, spin polarized or spin-valley-locked, only the valley polarized ground state is shown here for clarity. The gray criss-cross hatching indicates momentum-non-diagonal contribution of the corresponding order parameter. The dotted hatching corresponds to regions of the phase diagram with total momentum-non-diagonal contribution larger than 1 (using the same units as in the color scale) and an energy gap larger than 4 meV. These regions host a full WC (F-WC), a half WC (H-WC) and a quarter WC (Q-WC). The regions hatched with circles have a total momentum-non-diagonal contribution of at least 1 but no energy gap, this region corresponds to a thee quarter WC (TQ-WC). The HM is degenerate and can be valley polarized, spin polarized or spin-valley-locked, only the valley polarized ground state is shown here for clarity. \textbf{a)}, \textbf{b)}, \textbf{c)} were obtained with a reference projector that corresponds to populating half of the states in the valence band. \textbf{d)}, \textbf{f)}, \textbf{g)} were obtained with a reference projector that corresponds to entirely populating the valence band.}
    \label{fig:multiple_reference_WC}
\end{figure}

\newpage
\clearpage
\section{Hartree-Fock band structure}
\label{appx:Hartree-Fock band structure}

\begin{figure}[h!]
    \centering
     \includegraphics[width=12cm]{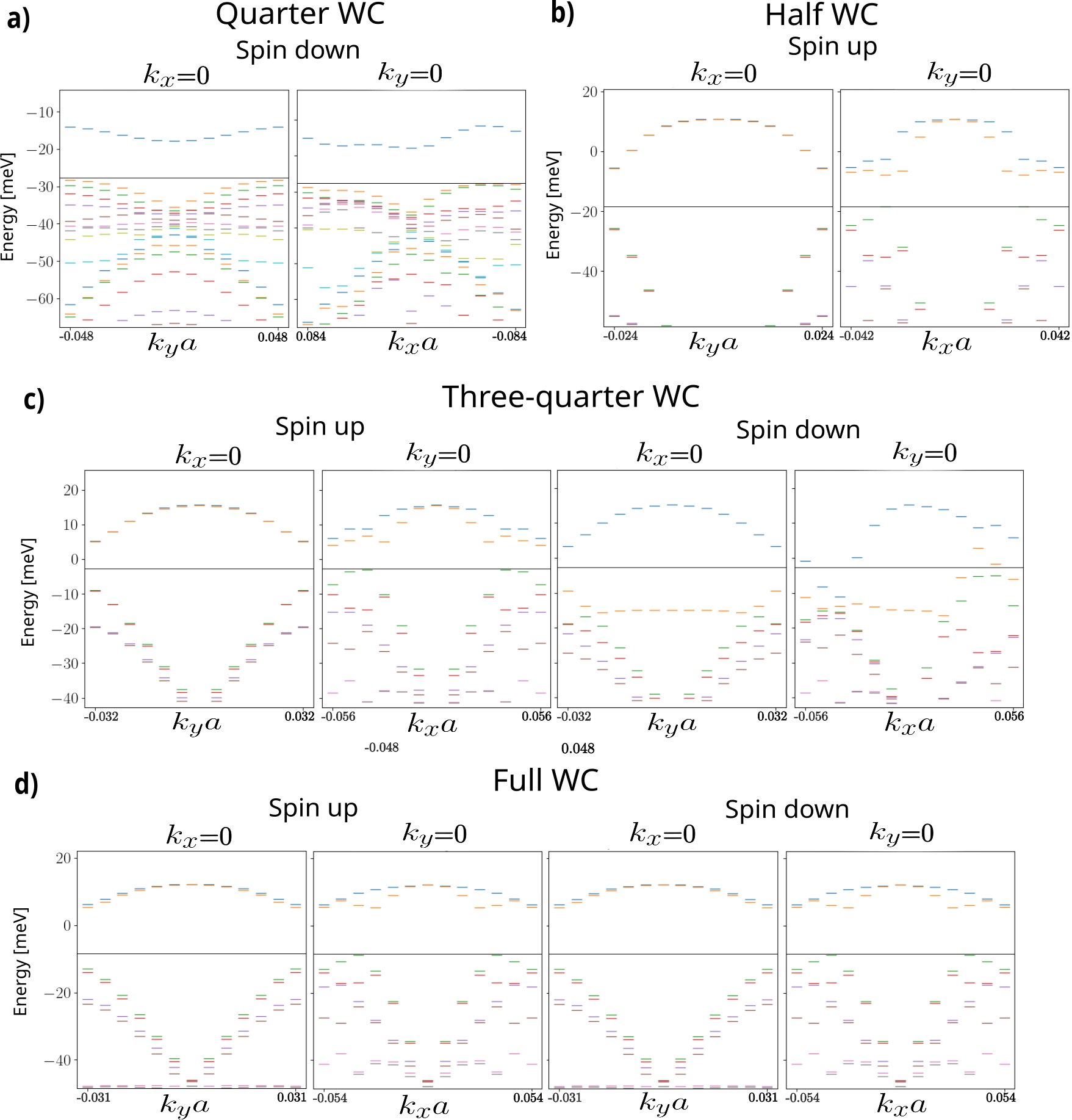}
     
    \caption{Cuts at $k_x=0$ and $k_y=0$ of the Hartree-Fock band structure. \textbf{a)}, \textbf{b)}, \textbf{c)}  and \textbf{d)} correspond to the same points in the phase diagram as panels 4, 1, 3 and 2 in figure \ref{fig:Real_space}, respectively. Only an excerpt of the bands with highest energy is shown. Since the half WC and quarter WC are fully spin polarized we only show the band structure corresponding to a single spin in \textbf{a)} and \textbf{b)}. For the three-quarter WC shown in \textbf{c)} there is an energy gap only in the spin up channel.}
    \label{fig:band structure}
\end{figure}




\end{appendix}

\end{document}